# A system to test 2D optoelectronic devices in high vacuum

Qinghua Zhao[1, 2, 3], Felix Carrascoso[3], Patricia Gant[3], Tao Wang[1,2], Riccardo Frisenda[3, *] and Andres Castellanos-Gomez[3, *]
[1]State Key Laboratory of Solidification Processing, Northwestern Polytechnical University, Xi'an, P. R. China.
[2]Key Laboratory of Radiation Detection Materials and Devices, Ministry of Industry and Information Technology, Xi'an, P. R. China
[3]Materials Science Factory. Instituto de Ciencia de Materiales de Madrid (ICMM-CSIC), Madrid, Spain.
riccardo.frisenda@csic.es , andres.castellanos@csic.es

The exploration of electronic and optoelectronic properties of two-dimensional (2D) materials has become one of the most attractive line of research since the isolation of graphene. Such '*all-surface materials*' present a strong sensitivity to environmental conditions and thus characterization of the devices based on these materials usually requires measurement systems operating in high-vacuum. However, conventional optoelectronic probe-station testing systems are are not compatible with high vacuum operation and vacuum-compatible versions are rather expensive. Here, we present a high-vacuum system specifically designed to test electronic and optoelectronic devices based on 2D materials. This system can be implemented with low budget and it is mostly based on the assembly of commercially available standard vacuum and optic components. Despite the simplicity of this system we demonstrate full capabilities to characterize optoelectronic devices in a broad range of wavelengths with fast pumping/venting speed and possibility of modulating the device temperature (room temperature to ~ 150°C).

Since the isolation of graphene and other 2D materials,[1-3] the scientific community has worked to apply these nanomaterials in several electronic and optoelectronic devices.[4, 5] Being ‘*all-surface materials’*, the properties of 2D materials are highly sensitive to the environmental conditions.[6-11] This has made it necessary the use of experimental testing systems that allow for a control of the atmospheric condition.[12, 13] In many cases, high vacuum is needed to get rid of the adsorbed moisture layer at the surface.[10] Conventional optoelectronic testing tools such as probe stations are not compatible with high vacuum operation and vacuum-compatible versions are rather expensive.

In this work we present a compact and inexpensive high-vacuum system to characterize electronic and optoelectronic devices based on 2D materials. The system is mainly based on the assembly of standard high-vacuum elements, requiring only few home-built parts. We supplement the system with probes that facilitate the electrical connection to the devices, and we employ fiber optic-based illumination to study the optoelectronic properties of the devices. We show that the system can be evacuated quickly to reach high-level vacuum conditions and that a fast control of the device temperature can be realized. We illustrate the operation of the system by characterizing the performance of Au-InSe-Au devices in dark and under external controlled illumination in high vacuum conditions.

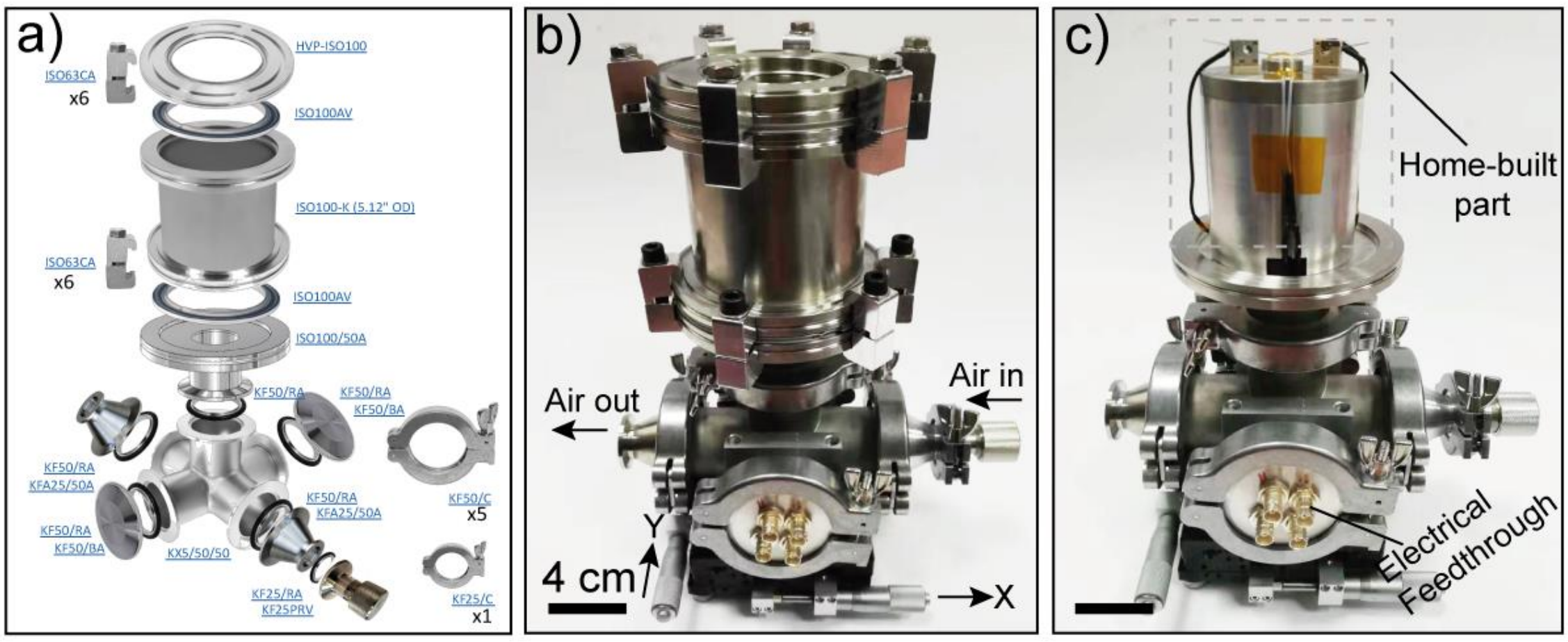

**Figure 1.** (a) Scheme of the commercial high-vacuum parts used for vacuum chamber building. (b, c) Photographs with (b) and without (c) the cap of the assembled vacuum chamber mounted onto a XY stage. The home-built parts are highlighted in panel (c).

**Table 1.** Summary of the components needed for the assembly of the small vacuum chamber.

| Vendor | Quantity | Part number | Description | Unit price |
|---|---|---|---|---|
| **Hositrad** | 1 | ISO100/50A | ISO-K 100 to NW50KF adaptor | 97.00 |
| | 2 | ISO100AV | ISO-K 100 centering ring | 16.50 |
| | 12 | ISO63CA | ISO-K 100 clamps | 1.90 |
| | 1 | KX5/50/50 | NW50KF 5-way cross | 240.00 |
| | 5 | KF50/RA | NW50KF centering ring | 5.00 |
| | 5 | KF50/C | NW50KF clamps | 6.00 |
| | 2 | KFA25/50A | NW25KF to NW50KF (reducer) | 18.00 |
| | 2 | KF50/BA | NW50KF blanks | 6.00 |
| | 1 | HVP-ISO100 | NW100 ISO-K viewport | 195.00 |
| | 1 | KF25/RA | NW25KF centering ring | 2.25 |
| | 1 | KF25/C | NW25KF clamps | 2.50 |
| **USA lab** | 1 | KF25PRV | KF25 Venting Valve | 59.00 |
| **Lesker** | 1 | ISO100-K (5.12" OD) | Full Nipple, Clamp Style | 138.70 |
| **RS** | 7 | 295-7942 | Hermetic BNC connector | 7.97 |
| **Optics-Focus** | 1 | MAXY-125L-25 | XY placement platform | 180.00 |
| **Linkam** | 3 | --- | Magnetic probe-tip holder | 280 |
| | | | **Total:** | 1967.78 € |

**Table 2.** Summary of the components needed for the assembly of the optical system used for inspection/illumination of the devices.

| Vendor | Quantity | Part number | Description | Unit price |
|---|---|---|---|---|
| AMScope | 1 | SKU: SA-HG-2 | Solid Aluminum Single-arm Boom Stand | 202.99 |
| Aliexpress | 1 | ZOOM_LENS | 400x zoom lens with coaxial illuminator | 185.99 |
| | 1 | DIGI_CAM | 21 MPix HDMI + USB camera | 77.88 |
| Thorlabs | 1 | M530F2 | High-power fiber coupled LED source (λ = 530 nm, other wavelengths available) | 371.82 |
| | 1 | LEDD1B | LED driver | 294.11 |
| | 1 | M28L01 | Multimode optical fiber (core 400 µm, other core sizes available) | 84.59 |
| | 1 | SCP05 | Miniature XY translator stage for the optical fiber (see Figure 5b3) | 146.49 |
| | 1 | SM05SMA | Adapter from optical fiber to miniature XY stage (see Figure 5b3) | 27.54 |
| | | | **Total:** | 1391.41 € |

**Table 3.** Summary of the auxiliary equipment used here to pump down the chamber and to perform the electrical measurements.

| Vendor | Quantity | Model | Description | Unit price |
|---|---|---|---|---|
| Keithley | 1 | Keithley 2450 | Source measurement unit for the electrical measurements of devices | 5250.00 |
| Edwards | 1 | T-Station 85H Dry NW40 | Turbo pump station with wide range pressure gauge | 6700.00 |
| | | | **Total:** | 11 950.00 € |

Figure 1a shows a scheme of the high vacuum parts (with model information) needed to assemble the vacuum chamber and 1b shows the actual photograph of the assembled setup fixed on a manual X-Y placement platform with a travel range of 30 mm × 30 mm and an adjustment accuracy of 1 μm. Two KFA25/50A reducers at the bottom of the chamber are used for the connection of molecular turbo pumping station (indicated as "air out" in the schematics) and venting valve ("air in"), which allows fast chamber atmosphere switching between vacuum and air. Figure 1c shows the interior of the chamber after removing the cap. A machined cylinder piece (marked by squared gray dash line in Figure 1c) is placed inside the chamber in order to reduce the pumping volume and to place the sample closer to the system optical window. Although a smaller chamber size would have led to even faster pumping speed, we choose to base our design in an ISO-100 standard size in order to facilitate the sample and probe placement operations. In the last years we have tested smaller chamber sizes finding them quite uncomfortable for the operator. The electrical feedthrough (homebuilt, see more details below) allows connecting the electronic components to perform the electrical measurements: source measure unit for electrical characterization of devices, programmable benchtop power supply to control the heater mounted on sample stage and multi-meter to test the thermistor resistance.

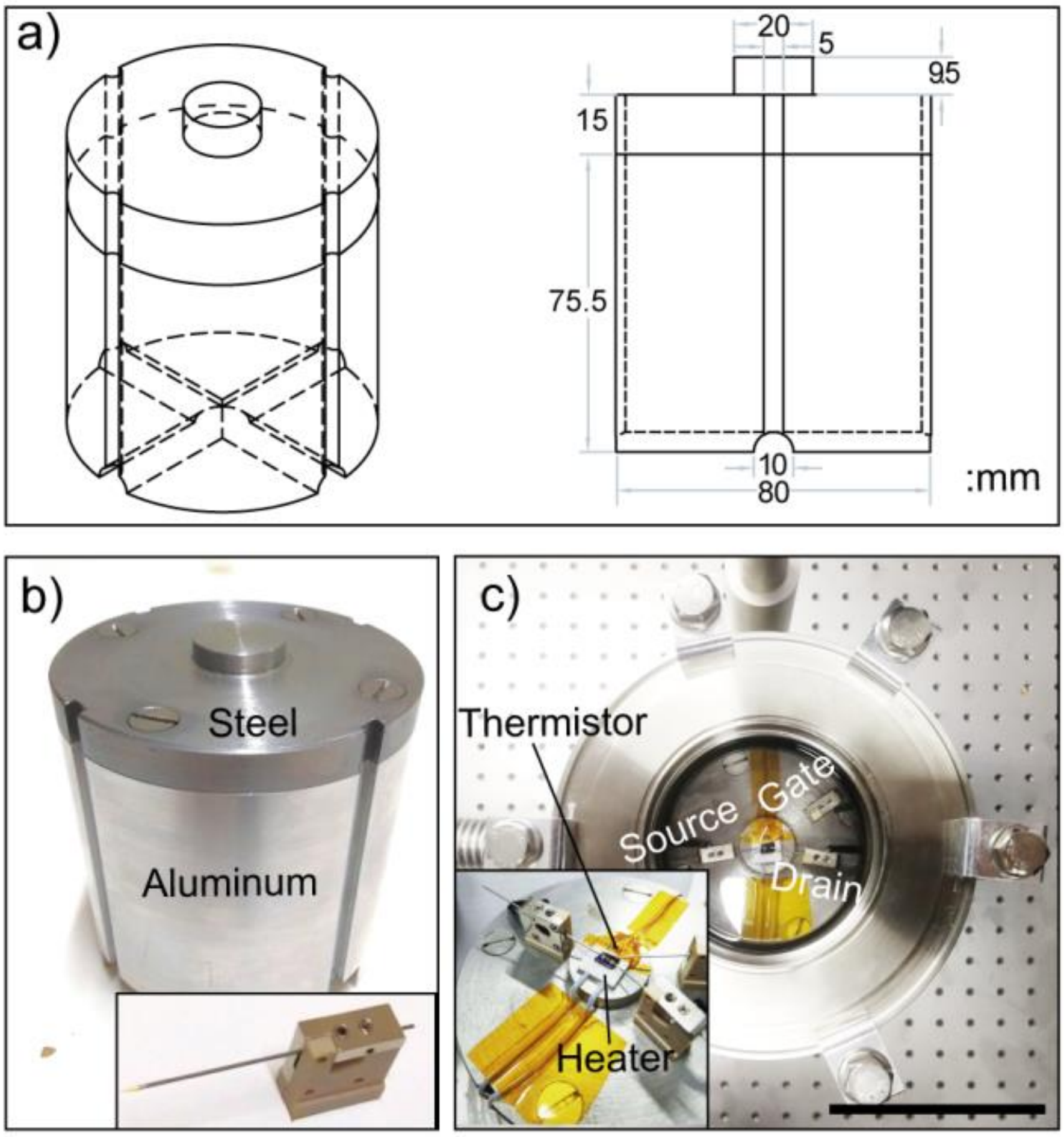

**Figure 2.** Home-built sample stage. (a) Isometric 3D sketch of home-built sample stage (left panel) and side view with indicated the corresponding dimensions in millimeters (right panel). (b) Close-up photographs of the as-machined home-built sample stage and the miniature probe with bottom magnets. (c) Top view of the sample stage surface as seen through the optical window. The inset is a close-up image of the ceramic heater, thermistor and three miniature probes (source, drain and gate). The scale bar is 50 mm.

The details of home-built parts necessary to complete the vacuum chamber system are shown in Figure 2 and Figure 3, respectively. Figure 2a shows a sketch of the home-built sample stage that sits inside the vacuum chamber. The corresponding dimensions are marked in the left panel of Figure 2a. Figure 2b shows a close-up picture of the home-built sample stage cylinder which consists of the top (magnetic) steel part and the bottom supporting aluminum cylinder part. The top steel part surface has been supplemented with a ceramic heater plate to heat up the devices (up to 200ºC), a thermistor used for temperature sensing (note that the thermistor used here limits the maximum operation temperature to 150ºC). Three miniature probes (Linkam) attached with magnets are used to contact the electrical pads of the devices to establish the source, drain and gate connections. The heater, the thermistor and the three probes are connected to a home-built BNC electrical feedthrough through coaxial cable (RG 178, BELDEN). An example of such an electrical feedthrough fabrication process is shown in Figure 3. Breifly, two holes are drilled on a blank KF50 flange and hermetic BNC nut-jam connectors are screwed to ensure a vacuum tight sealing.

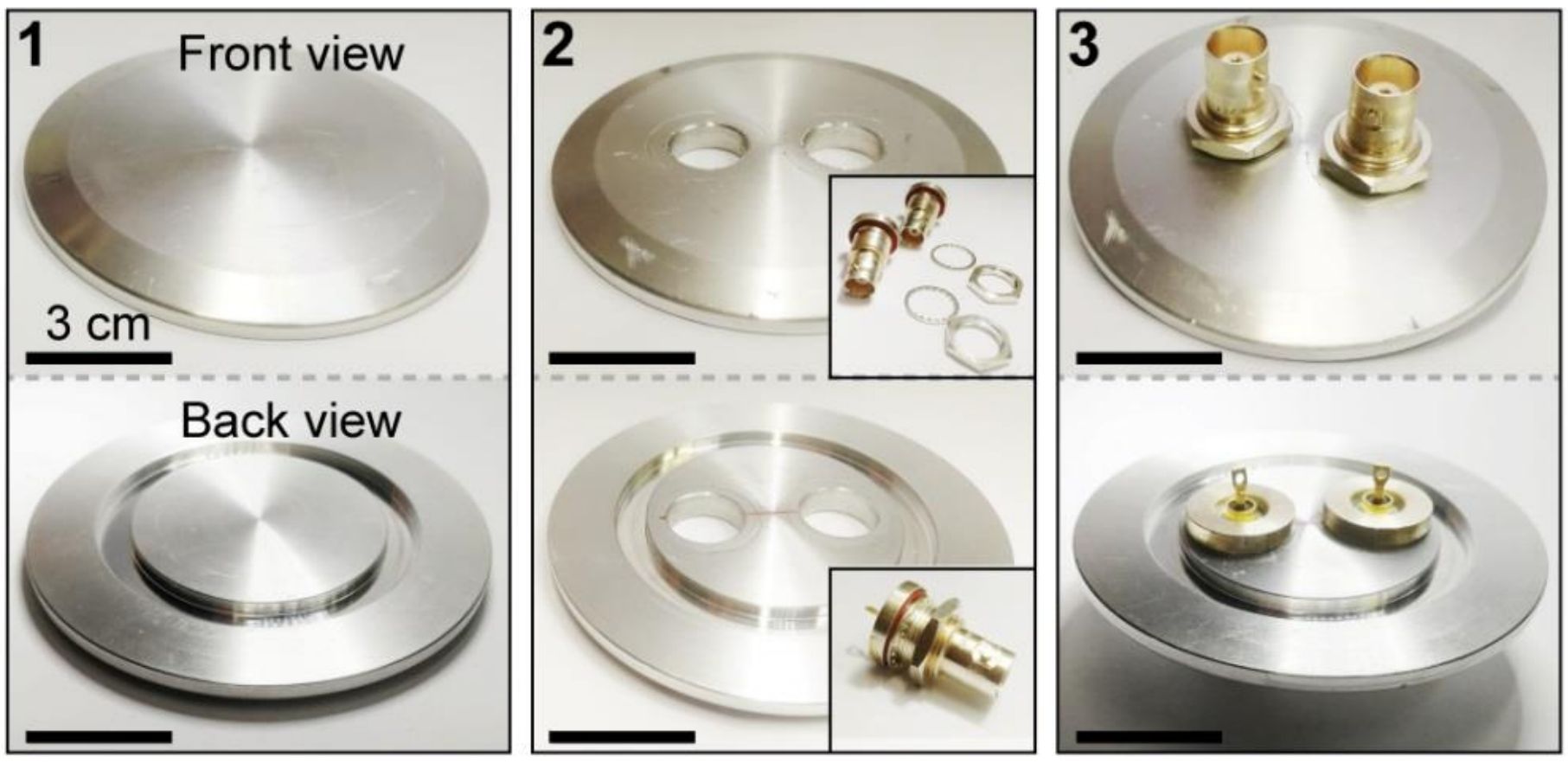

**Figure 3.** Fabrication process of an electrical feedthrough. The front view and back view of a KF50 flange (1), the KF50 flange with two drilled holes (2) and two hermetic BNC nut-jam connectors are screwed on the KF50 flange (3).

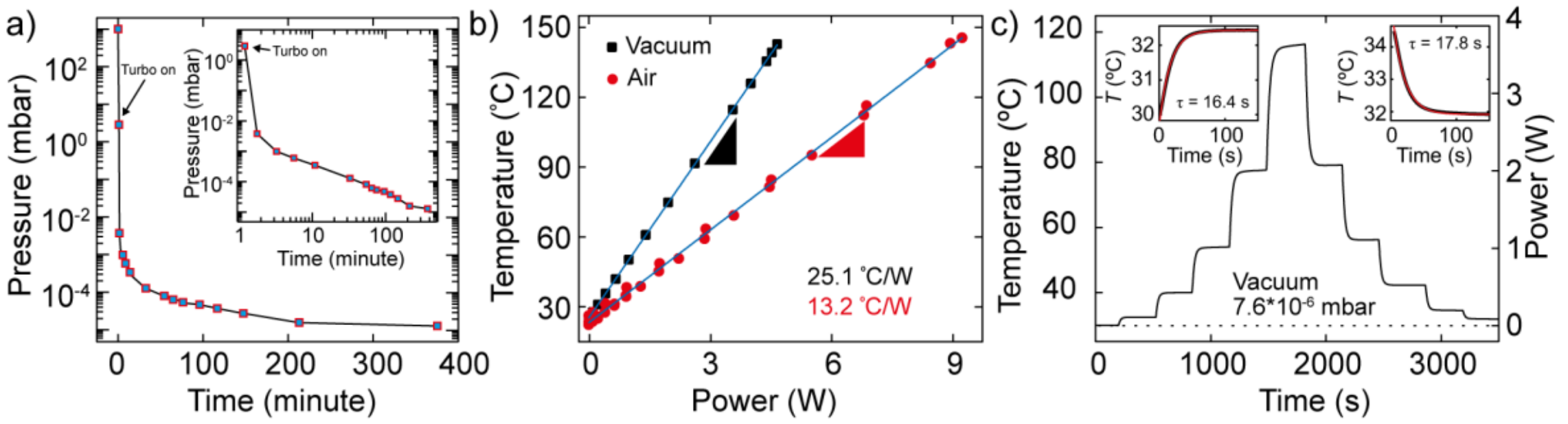

**Figure 4.** Pressure and temperature calibration of the vacuum chamber. (a) Recorded pressure in the chamber as a function of molecular turbo pumping time. The inset is the plot of pressure versus time in log-log scale. (b) Relationship between surface temperature and working power of heater measured under atmosphere and vacuum (~ $10^{-5}$ mbar) condition. (c) Continuous temperature control recorded in time as a function of working power of heater under high-vacuum ($7.6 \cdot 10^{-6}$ mbar) condition. The insets are exponential growth/decay fits of the heating (left) and cooling (right) temperature as a function of time.

Figure 4a shows the internal pressure of the chamber as a function of time when connected to a molecular turbo pumping station (Edwards, T-station 85H). Due to the reduced pumping volume the chamber reaches low enough pressure in relatively short period of time. The internal pressure of chamber can be reduced from atmosphere pressure to ~$10^{-4}$ mbar within less than ~30 minutes of pumping time. Note that upon pumping over long periods of time (typically 10 hours overnight) we have regularly achieved pressures lower than $10^{-5}$ mbar. Figure 4b shows the calibration of the ceramic heater, the temperature values are recorded from room temperature up to ~150 ℃ in ambient (red circles) and in vacuum (black squares) conditions by controlling the heater working power. The good agreement between the experimental data and the linear fit (blue line) confirm that the heater temperature depends linearly on the heater working power. From the linear fits we find the heater calibration values 13.2 ℃/W in air and 25.1 ℃/W in vacuum. To test the temperature response time of the system we recorded the temperature as a function of time while the heater working power is changed in increasingly large steps, from 0 W to 4 W, every 600 seconds. From Figure 4c, one can observe that the stable temperature levels are always accompanied by the fast temperature raising and dropping while turning up or turning down the heater power. Details on the heating and cooling temperature change are shown in left and right inset panel in Figure 4c. Both of the processes are well fitted to a exponential growth/decay function with the time parameters $\tau \approx 16.4\text{-}17.8$ s.

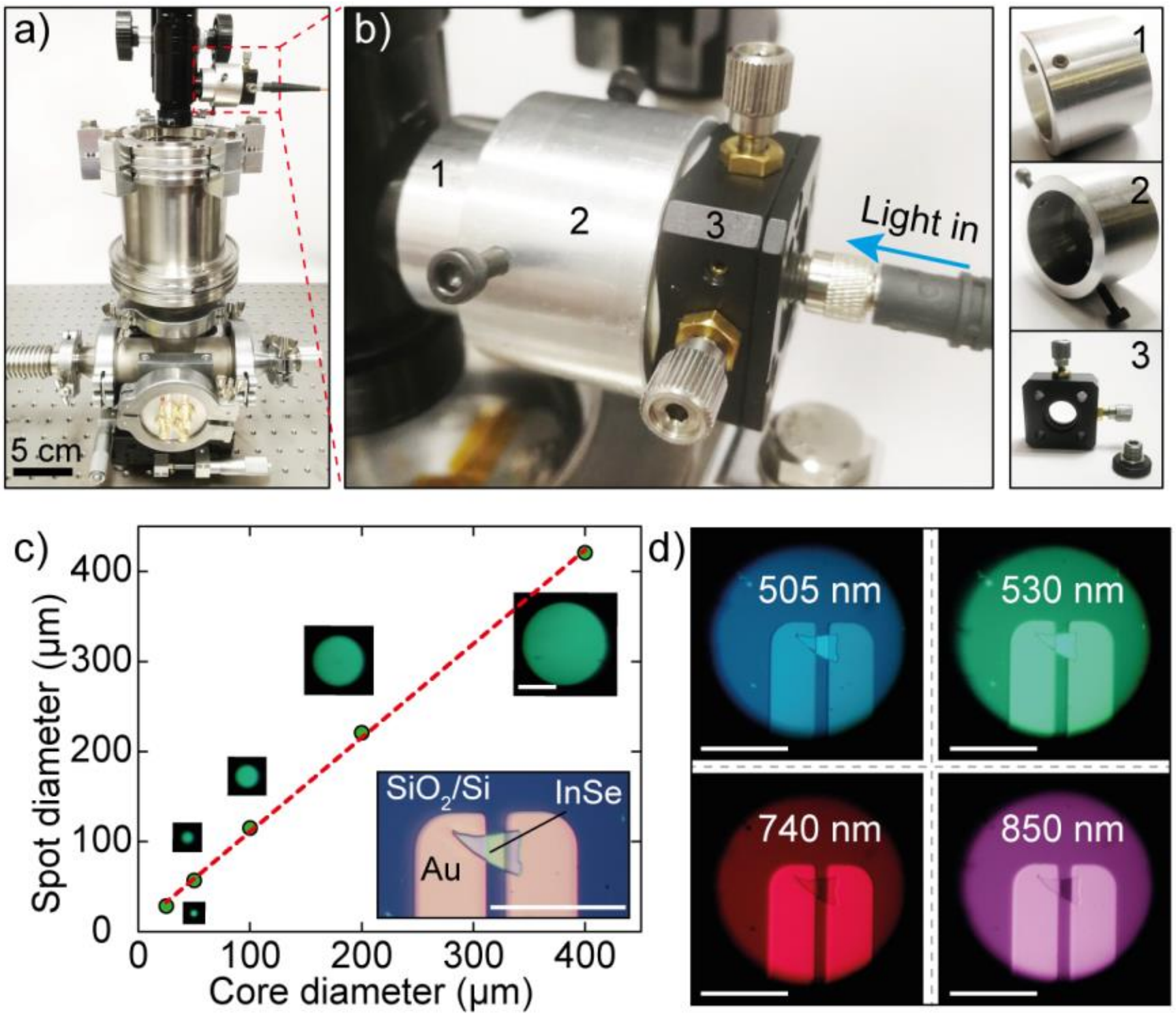

**Figure 5.** (a) Optical image of the optoelectronic measurement system. (b) Close up pictures of a home-made fiber holder, highlighting the three components separately. (c) Calibration of the illumination spot size as a function of core size of the optical fiber with a fixed zoom lens magnification. The inset is a Au-InSe-Au device under global illumination with a white LED source. (d) Photographs of the Au-InSe-Au device being illuminated respectively with 505 nm, 530 nm, 740 nm and 850 nm LED illuminators by projecting the core of a 400 μm fiber onto the sample surface. All the scale bars in panel (c) and (d) are 200 μm.

In order to study the optoelectronic response of devices we place the vacuum chamber under a long working distance zoom lens that allows one both imaging the devices and shining light on them. Figure 5a shows an optical image after mounting the chamber under the zoom lens imaging/illumination system. In order to photo-excite the devices with a controlled illumination, the coaxial zoom lens system can be supplemented with a home-built part (highlighted with a red dashed square in Figure 5a) that holds a multimode optical fiber. The details about the home-made fiber holder are shown by a close-up picture in Figure 5b and the fabrication details can be found in Figure S1. The three isolated components are shown in the right panels. The fiber holder allows one to place the fiber core on the image plane of the zoom lens system, thus projecting an image of the fiber core on the sample that produces a circular spot with uniform power density on the sample surface. Note that the size of the circular spot is both controlled by the optical fiber core size and the magnification of the zoom lens system. Figure 5c shows a comparison of the spots acquired by connecting 5 optical fibers with different core sizes ranging from 25 μm to 400 μm. A linear relationship between the projected illumination spot size (on a 285 nm $SiO_2$/Si substrate) and the optical fiber core diameter (for a fixed magnification of the zoom lens) is obtained. The projection of the fiber core on the sample in combination with high-power fiber coupled LED sources (Thorlabs) yields highly homogenous and speckle free spots that result very convenient to determine the figures of merit of a photodetector, like photoresponsivity and detectivity. We address the reader to the Supporting Information (Figure S2) for a comparison between the intensity profile obtained with this illumination method with respect to a Gaussian-shape laser spot. Moreover, this illumination method allows for easily change of the illumination wavelength without a substantial change in the spot size and shape. Figure 5d shows a Au-InSe-Au device illuminated by projecting the core of a 400 μm fiber coupled LED illuminators with 505 nm - 850 nm of central wavelength.

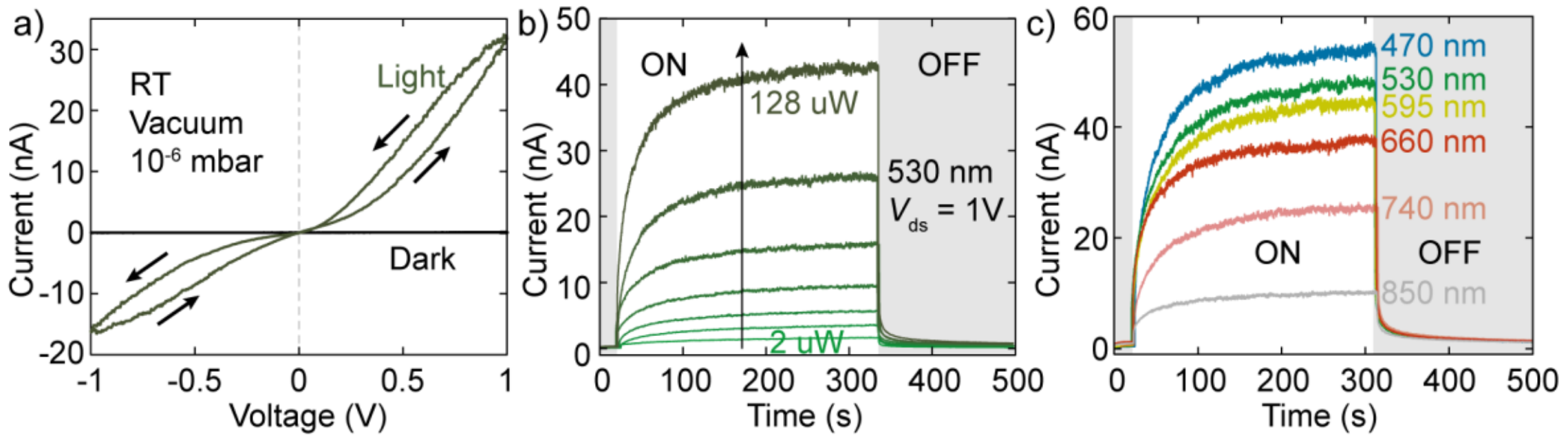

**Figure 6.** Optoelectronic characterizations of a Au-InSe-Au photodetector in vacuum at room temperature. (a) Current-voltage characteristics of a pristine Au-InSe-Au photodetector recorded in dark and under 530 nm LED illumination. (b, c) Current recorded at the $V_{ds}$ = 1V as a function of time while switching ON and OFF the 530 nm LED illuminator with different power (b) and with the LEDs of different wavelength (470 nm to 850 nm) with a fixed power of 100 μW (c).

In the following, we illustrate the operation of the system by characterizing the optoelectronic properties of the Au-InSe-Au photodetector (shown in the inset in Figure 5c), assembled by deterministic placement of an InSe flake onto pre-patterned gold electrodes,[14-16] under high vacuum condition at room temperature. We use a source measurement unit (Keithley 2450) to perform electrical measurements on the device in the dark state and upon illumination. Current-voltage (*I*-*V*) and current-time (*I*-*t*) curves are typical characteristics for measuring optoelectronic properties of photodetectors. Figure 6a shows the source-drain current as a function of voltage of a pristine Au-InSe-Au photodetector recorded under dark condition and illumination of 530 nm LED illuminator. Negligible current flows through the device when kept in dark condition, while an obvious increase in current can be observed when the device was measured under illumination due to the photogeneration of charge carriers in the device.[17] The current hysteresis observed in the light current-voltage curve suggests the existence of traps in pristine InSe crystals under vacuum condition.[8] The response of the device to light is then studied by measuring the source-drain current flowing through the device at the fixed bias voltage as a function of time while the illumination is switched ON and OFF. Figure 6b shows an example of that kind of measurement that allows determining both the photocurrent (the difference between the current with light ON and light OFF) and the response time of the device. The response time of a photodetector is usually determined by measuring the time that it takes to go from the 10% to the 90% (or from the 90% to the 10%) of the generated photocurrent under modulated excitation illumination, either in the rising or falling edge.[17] By repeating the measurements for different illumination wavelength at same power one can obtain the photocurrent spectrum of the device.

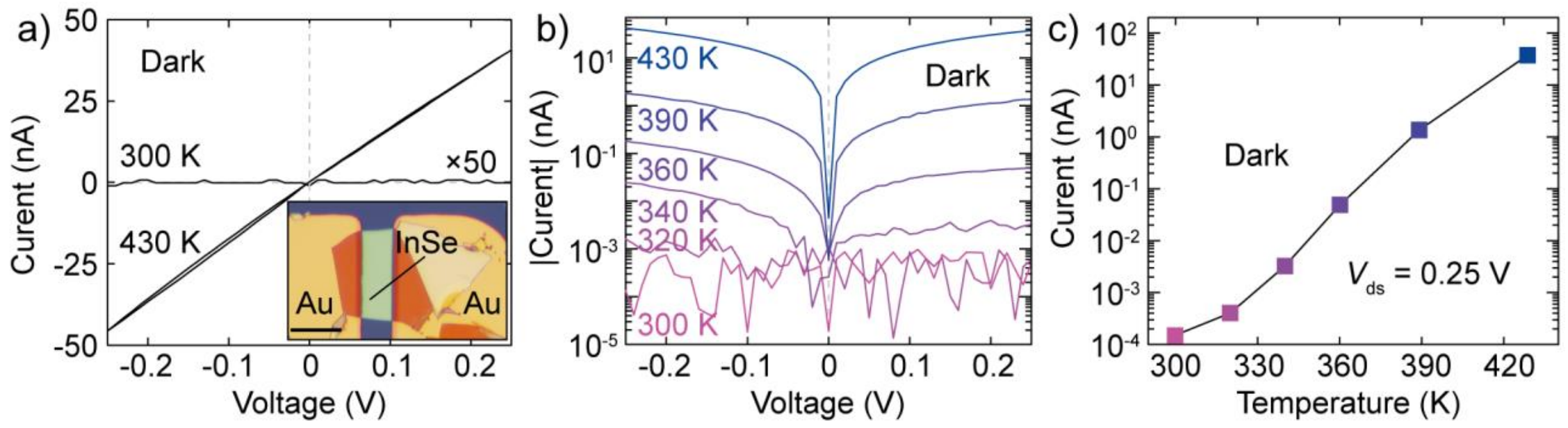

**Figure 7.** Temperature-dependent characterization of a Au-InSe-Au photodetector in vacuum. (a) Current-voltage characteristics of a Au-InSe-Au photodetector recorded at 300 K and 430 K in dark. The inset in panel (a) shows an image of the Au-InSe-Au photodetector. (b) Current-voltage characteristics recorded in dark as a function of temperature. (d) Current as a function of temperature recorded in dark at $V_{ds}$ = 0.25 V.

In order to demonstrate the temperature control of the setup, we carry out a set of optoelectronic characterization of an Au-InSe-Au device by varying the testing temperature. Figure 7a shows a linear plot of current-voltage characteristics of the Au-InSe-Au photodetector plotted recorded at 300 K and 430 K in dark condition. Noise-level current flows through the device in dark at 300 K while the current increases noticeably when the temperature is increased to 430 K. The temperature dependency of the *I*-*V* curves of the device recorded in dark with semi-logarithmic scale are reported in Figure 7b. The *I*-*V*s recorded while increasing the temperature from $T$ = 300 K to $T$ = 430 K show a current increase both at positive and negative bias voltages and an increase of the symmetry. Based on the *I*-*V*s recorded in dark, we extract the current value at $V_{ds}$ = 0.25 V and plot them as a function of temperature in Figure 7c. The dark current flowing through the device increases with the temperature in an exponential way (that appears linear in the semi-logarithmic representation of Fig. 7c), this could be due to an increase of carrier density by thermal excitation in the InSe or to a lowering of the contact resistance.[18-20]

## Conclusions

In summary, we presented an inexpensive system to characterize optoelectronic devices based on 2D materials in high-vacuum. The whole system can be assembled based on commercially available standard vacuum elements and only few home-built parts are needed. We demonstrate that this system can reach a pressure <$10^{-5}$ mbar and it allows for a fast testing temperature modulation from room temperature to ~ 150 °C. We show how the system can be used to characterize devices based on 2D materials. In fact, we illustrate the operation of our system by testing the electrical and optoelectronic characteristics of a InSe device at a pressure of $10^{-6}$ mbar with different illumination powers (ranging from 2 μW to 128 μW), with different wavelengths (from 470 nm to 850 nm) and at different temperatures (RT to 130ºC). This setup can be an alternative to more expensive commercially available systems and thus we believe that such kind of system can be easily implemented in labs with low budget and satisfy the characterization requirements for 2D materials optoelectronic devices.

**Acknowledgements**

This project has received funding from the European Research Council (ERC) under the European Union's Horizon 2020 research and innovation programme (grant agreement n° 755655, ERC-StG 2017 project 2D-TOPSENSE). R.F. acknowledges the support from the Spanish Ministry of Economy, Industry and Competitiveness through a Juan de la Cierva-formación fellowship 2017 FJCI-2017-32919. QHZ acknowledges the grant from China Scholarship Council (CSC) under No. 201700290035.

# Supporting Information:

# A system to test 2D optoelectronic devices in high vacuum

Qinghua Zhao[1, 2, 3], Felix Carrascoso[3], Patricia Gant[3], Tao Wang[1,2], Riccardo Frisenda[3, *] and Andres Castellanos-Gomez[3, *]

[1]State Key Laboratory of Solidification Processing, Northwestern Polytechnical University, Xi'an, P. R. China.
[2]Key Laboratory of Radiation Detection Materials and Devices, Ministry of Industry and Information Technology, Xi'an, P. R. China
[3]Materials Science Factory. Instituto de Ciencia de Materiales de Madrid (ICMM-CSIC), Madrid, Spain.
riccardo.frisenda@csic.es , andres.castellanos@csic.es

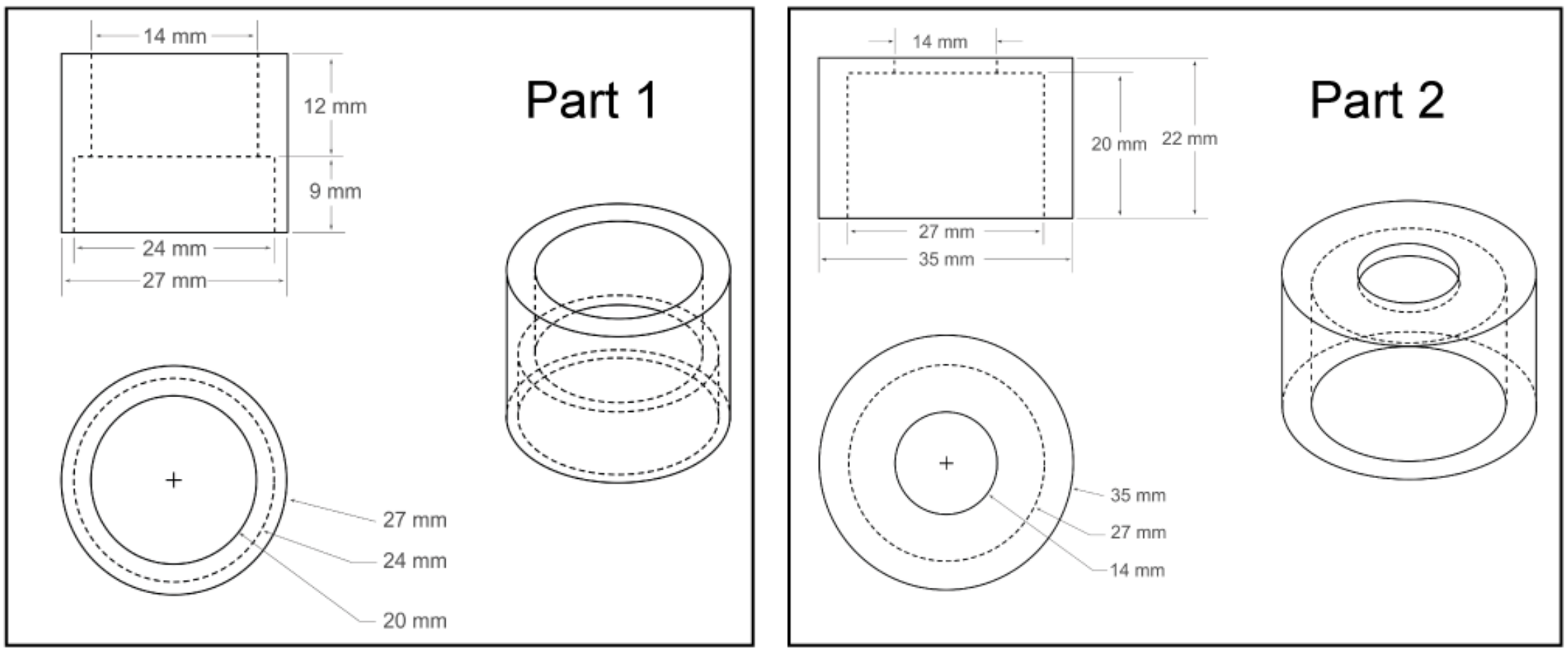

**Figure S1:** The fabrication dimension of the home-made parts (highlighted by "1" and "2" in Figure 5b) of the fiber holder.

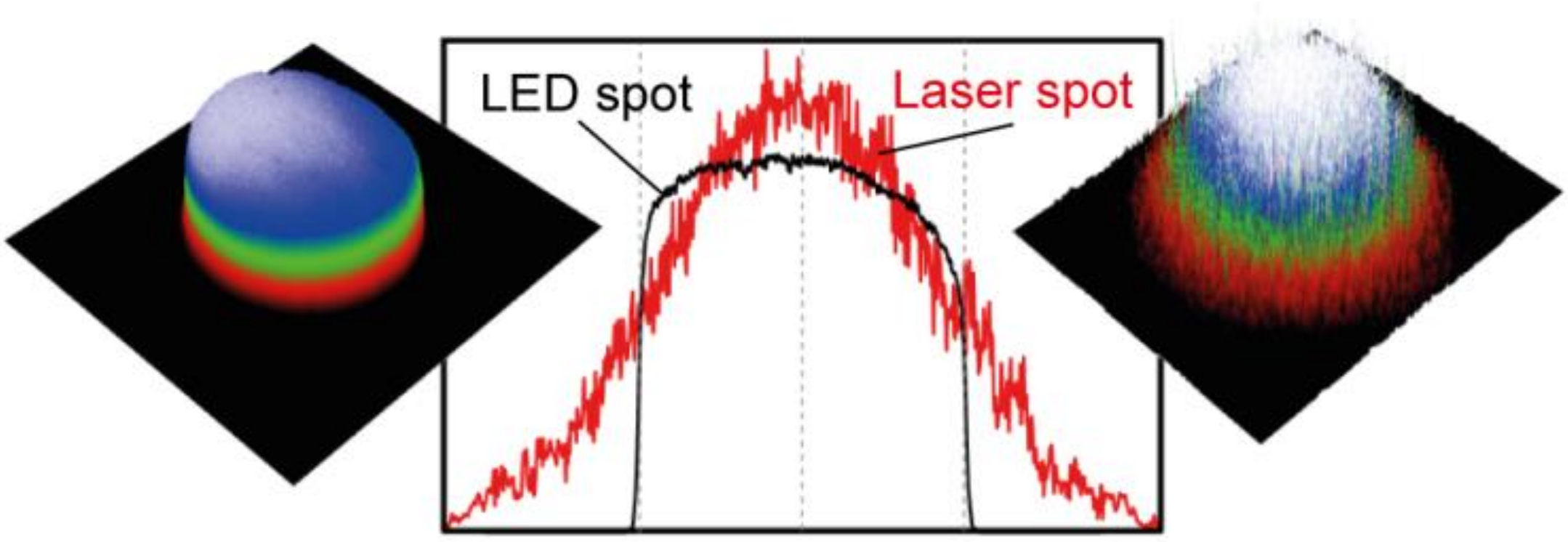

**Figure S2:** The comparison between the intensity profile obtained with LED illumination method with respect to a Gaussian-shape laser spot.